\documentclass[12pt]{iopart}

\begin{document}
\title[Swendsen-Wang and Wolff Algorithms]
{Dynamical Critical Behaviors of the Ising Spin Chain: Swendsen-Wang and
  Wolff Algorithms}    
\author{P.~L.~Krapivsky}
\address{Center for Polymer Studies 
and Department of Physics, Boston University, Boston, MA 02215, USA}
\ead{paulk@bu.edu}  

\begin{abstract}
  
  We study the zero-temperature Ising chain evolving according to the
  Swendsen-Wang dynamics. We determine analytically the domain length
  distribution and various ``historical'' characteristics, e.g., the density
  of unreacted domains is shown to scale with the average domain length as
  $\langle l\rangle^{-\delta}$ with $\delta=3/2$ (for the $q$-state Potts
  model, $\delta=1+q^{-1}$).  We also compute the domain length distribution
  for the Ising chain endowed with the zero-temperature Wolff dynamics.

\end{abstract}  

\pacs{02.50Ey, 05.40.+j, 82.20.Mj}

%\maketitle

\section{Introduction}

Interesting and relatively poorly understood dynamical critical behaviors
occur when statistical-mechanical systems are quenched from a disordered
phase to their critical points.  For the Ising spin systems, two very popular
dynamics were introduced long ago by Glauber \cite{glauber} and Kawasaki
\cite{kawasaki}.  Glauber and Kawasaki algorithms are the simplest dynamical
rules based on local moves --- single spin flips for the non-conservative
Glauber dynamics and spin exchanges for the conservative Kawasaki dynamics.
Glauber's and other single spin-flip dynamics, particularly the Metropolis
algorithm, have become a powerful tool for understanding the equilibrium
behavior of the statistical-mechanical systems well away from the critical
temperature \cite{NB,LB}. The simulation becomes very slow at criticality,
however, and to overcome this difficulty non-local moves, such as cluster
flips, have been suggested.  The Swendsen-Wang \cite{sw} and Wolff \cite{w}
dynamics are two well-known cluster algorithms that are widely used in
elucidating the equilibrium critical behavior in statistical physics and
lattice field theory.  The dynamical critical behavior of these algorithms is
an active area of research, see \cite{S} and references therein.  This work
heavily relies on simulations, e.g., the value of the dynamical critical
exponent for the two-dimensional Ising model endowed with the Swendsen-Wang
dynamics is known only numerically; analytic studies of the Swendsen-Wang
dynamics have been limited so far to the Ising model on the complete graph,
that is to the Curie-Weiss or the mean-field model \cite{klein,domany,CF}.

The purpose of this paper is to investigate the dynamical aspects of the
cluster algorithms --- particularly the Swendsen-Wang algorithm --- in the
simplest possible setting, that is in one dimension. The critical temperature
is usually $T_c=0$ for one-dimensional systems. The zero-temperature dynamics
can be quite peculiar, for example the Ising spin chain subject to the
zero-temperature Kawasaki dynamics freezes \cite{cornell,1d}.
Non-conservative dynamics, however, usually bring the system to the ground
state, e.g., the zero-temperature Ising-Glauber chain reaches the ground
state in a time $\tau\sim {\cal L}^2$ (here ${\cal L}$ is the system size),
that is the dynamical exponent is $z=2$ for the Glauber algorithm
\cite{glauber,bray}. We will see that $z=0$ for the Swendsen-Wang algorithm
in one dimension; more precisely, $\tau\sim \ln {\cal L}$.  The
one-dimensional version of the Swendsen-Wang algorithm is also an appealingly
simple model that is reminiscent of other important models of phase-ordering
dynamics like the time-dependent Ginzburg-Landau equation with no thermal
noise (i.e., $T=0$) \cite{NK,BDG}.  Furthermore, the one-dimensional
Swendsen-Wang algorithm provides a useful laboratory to probe not merely the
dynamical critical exponent but various much more subtle dynamical
characteristics.

This paper is organized as follows. In section 2 we show that the Ising chain
endowed with zero-temperature Swendsen-Wang dynamics exhibits scaling with
the average length growing exponentially with time.  Section 2 also contains
the derivation of the domain length distribution and a number of subtle
statistical properties of the domains like the density of domains which never
flipped. In section 3 we investigate the Ising chain endowed with
zero-temperature Wolff dynamics.  This model has been previously studied by
Derrida and Hakim \cite{DH}; here we further analyze the model, particularly
we determine the domain length distribution.  A summary is given in the last
section 4.

\section{Swendsen-Wang Dynamics}

In one dimension, the Ising spin chain can be thought as an array contiguous
alternating domains of up and down spins. At zero temperature, the
Swendsen-Wang dynamics amounts for randomly choosing a domain and flipping
it.  The flipping of a domain $I$ implies that it merges with its left and
right neighboring domains $I_-$ and $I_+$ thus forming a single domain
$I_-\cup I\cup I_+$.

\subsection{Domain Length Distribution}

Let $N_l(t)$ is the number of domains of length $l$ and $N(t)=\sum_{l\geq 1}
N_l(t)$ is the total number of domains.  The average number of domains that
flip in an infinitesimal time interval $\Delta t$ is equal to $N(t)\,\Delta
t$. In every flipping event three domains merge into a single one, so $N(t)$
changes according to
\begin{equation}
\label{Nt}
N(t+\Delta t)=N(t)-2\Delta t\, N(t).
\end{equation}
Similarly, $N_l(t)$ evolves according to
\begin{eqnarray}
\label{Nlt}
N_l(t+\Delta t)=N_l(t)-3\Delta t\,N_l(t)+\Delta t\sum_{i+j+k=l}
\frac{N_i(t)}{N(t)}\,N_j(t)\,\frac{N_k(t)}{N(t)}\,.
\end{eqnarray}  
The term $3\Delta t\,N_l(t)$ accounts for the loss that occurs when the
domain or either of its neighbors is flipped, while the last term on the
right-hand side of Eq.~(\ref{Nlt}) accounts for the gain due to the flipping
of a domain of length $j$ followed by immediate merging with two adjacent
domains of lengths $i$ and $k$.  Equation (\ref{Nt}) is obviously exact. The
linear loss term in (\ref{Nlt}) is also exact, while the non-linear gain term
is exact only if the sizes of adjacent domains are uncorrelated. However,
whenever a domain merges with two adjacent domains, the resulting domain does
not acquire any correlation with the neighbors, i.e., correlations are not
dynamically generated.  (See Ref.~\cite{BDG} for a detailed justification in
the context of a somewhat similar model, viz. the noiseless time-dependent
Ginzburg-Landau equation in which domains also merge with their neighbors.)
Therefore, if initially the domain lengths were uncorrelated, they remain
uncorrelated at all later times and Eq.~(\ref{Nlt}) is exact.

The total length of the system is ${\cal L}=\sum_{l\geq 1} lN_l(t)$.  In the
thermodynamic limit ${\cal L}\to\infty$, it is convenient to use the domain
length densities $n_l(t)=N_l(t)/{\cal L}$ and the (total) domain density
$n(t)=\sum_{l\geq 1} n_l(t)=N(t)/{\cal L}$. {}From (\ref{Nt}) we find that
the domain density evolves according to
\begin{equation}
\label{nt}
\frac{dn}{dt}=-2n.
\end{equation}
Solving (\ref{nt}) gives $n(t)=n(0)\,e^{-2t}$. Therefore the average domain
size $\langle l\rangle=1/n$ increases exponentially. In contrast, the average
size exhibits an algebraic growth $\langle l\rangle\sim t^{1/z}$ \cite{B} in
most models describing domain coarsening following a quench to zero
temperature.

Likewise, the equation for $N_l(t)$ leads to 
\begin{equation}
\label{nlt}
\frac{dn_l}{dt}=-3n_l+\sum_{i+j+k=l}\frac{n_i n_j n_k}{n^2}\,. 
\end{equation}
The form of Eqs.~(\ref{nlt}) suggests to consider normalized densities
$\rho_l(t)=n_l(t)/n(t)$.  These quantities satisfy
\begin{equation}
\label{rlt}
\frac{d\rho_l}{dt}=-\rho_l+\sum_{i+j+k=l}\rho_i \rho_j \rho_k. 
\end{equation}
Equations (\ref{nlt})--(\ref{rlt}) are mathematically similar to equations
describing the 3-particle coalescence process \cite{K} and can be solved
accordingly. Introducing the generating function
\begin{equation}
\label{Rdef}
R(t,x)=\sum_{l\geq 1} x^l\rho_l(t)
\end{equation}
we convert an infinite system of equations (\ref{rlt}) into a single equation
\begin{equation}
\label{Rt}
\frac{\partial R}{\partial t}=-R+R^3\,.
\end{equation}
Solving (\ref{Rt}) gives
\begin{equation}
\label{Rzt}
R(t,x)=\frac{e^{-t}\,R_0(x)}{\sqrt{1-\left(1-e^{-2t}\right)R_0^2(x)}}
\end{equation}
with $R_0(x)\equiv R(0,x)$.  For instance, consider the evolution starting
from the highest energy antiferromagnetic state. In this case,
$\rho_l(0)=\delta_{l1}$, i.e. $R_0(x)=x$. Inserting this into (\ref{Rzt}) and
expanding in powers of $x$ we obtain $\rho_{2l}(t)\equiv 0$ and
\begin{equation}
\label{raf}
\rho_{2l+1}(t)=e^{-t}\left(\frac{1-e^{-2t}}{4}\right)^l {2l\choose l}\,.
\end{equation}
The original densities read
\begin{equation}
\label{naf}
n_{2l+1}(t)=e^{-3t}\left(\frac{1-e^{-2t}}{4}\right)^l {2l\choose l}\,.
\end{equation}
In the scaling limit $l,t\to\infty$ with the scaling variable
\begin{equation}
\label{scal}
L=l e^{-2t}={\rm finite},
\end{equation}
the densities become 
\begin{equation}
\label{nscal}
n_l(t)=e^{-4t}\,{\cal F}(L),\quad 
{\cal F}(L)=\frac{1}{\sqrt{2\pi L}}\, \exp(-L/2).
\end{equation}
For other initial conditions it is quite difficult to extract explicit
results for $\rho_l$ from the general solution (\ref{Rzt}) for the generating
function.  The most natural situation arises when the system at $T=\infty$ is
suddenly quenched to $T=0$.  The appropriate initial condition is completely
uncorrelated with each spin taking the values $\pm 1$ independently and with
equal probabilities.  Then $\rho_l(0)=2^{-l}$, or $R_0(x)=x/(2-x)$ thus
leading to a relatively simple exact expression for the generating function.
The resulting $\rho_l$ admits a neat expression
\begin{equation}
\label{random}
\rho_l(t)=\tau^l\,{}_2F_1\left(l,\frac{1}{2};1;
  \frac{1-2\tau}{1-\tau}\right),
\qquad \tau=\frac{1+\sqrt{1-e^{-2t}}}{2}\,.
\end{equation}
An apparent simplicity of this solution is illusory as (\ref{random})
involves the hypergeometric function.  Fortunately, the most interesting
scaling behavior (\ref{nscal}) is universal, i.e., independent on initial
conditions (modulo the assumption that the decay $\rho_l(0)$ vs. $l$ is
sufficiently steep).  Therefore in the following we focus on the
antiferromagnetic initial condition. 

Note that the domain length distribution $n_l(t)$ is beyond the reach of
analytical approaches \cite{B} for the majority of models of domain
coarsening; e.g., for the Ising chain endowed with zero-temperature Glauber
dynamics the distribution $n_l(t)$ is still unknown although a few exact and
approximate results are available \cite{DZ,KB}.  The Swendsen-Wang dynamics
is obviously more tractable than the Glauber dynamics --- we determined
$n_l(t)$ for the antiferromagnetic initial condition and an exact scaling
expression in the general case.

\subsection{Domain Number Distribution}

For the Swendsen-Wang dynamics it is also possible to probe analytically
various historical characteristics. The simplest such quantity is the density
$m_p(t)$ of domains composed of $p$ `parent' domains that never flipped.
(Each such domain can of course include domains that flipped during the time
interval $(0,t)$.)  The domain number distribution $m_p(t)$ is formally
defined as follows \cite{KB}.  Initially $m_p(0)=\delta_{p1}$.  In every
merging event, the central domain flips while the two adjacent domains do not
flip; therefore if they have $\alpha$ and $\beta$ parents, respectively, the
resulting domain has $\alpha+\beta$ parents.  The domain number distribution
$m_p(t)$ evolves as follows:
\begin{equation}
\label{mpt}
\frac{dm_p}{dt}=-3m_p+\sum_{\alpha+\beta=p}\frac{m_\alpha m_\beta}{n}\,. 
\end{equation}
A solution to these equations has an exponential form
\begin{equation}
\label{maa}
m_p=A\,a^{p-1}.
\end{equation}
This ansatz reduces an infinite system (\ref{mpt}) to a couple of
differential equations
\begin{equation}
\label{aa}
\frac{dA}{dt}=-3A,\qquad
\frac{da}{dt}=n^{-1}A,
\end{equation}
where $n(t)=e^{-2t}$, see Eq.~(\ref{nt}).  Solving (\ref{aa}) subject 
$A(0)=1, a(0)=0$ (implied by the initial condition $m_p(0)=\delta_{p1}$) we
finally obtain
\begin{equation}
\label{msol}
m_p(t)=e^{-3t}\left(1-e^{-t}\right)^{p-1}.
\end{equation}
Thus $P=p\,e^{-t}$ is the scaling variable and the 
scaling form is simply exponential: 
\begin{equation}
\label{msol-scal}
m_p(t)=e^{-3t}{\cal G}(P), \qquad {\cal G}(P)=e^{-P}.
\end{equation}

The survival of a domain with and without merging are characterized by the
domain persistence exponents $\psi$ and $\delta$ which were investigated for
the Ising-Glauber \cite{KB} and Ising-Kawasaki \cite{BK} spin chains. (Domain
persistence and closely related cluster persistence exponents were also
studied for several other models \cite{LM,DLM,HA,R}.)  The exponent $\delta$
describes the decay of primordial domains
\begin{equation}
\label{delta}
m_1\sim \langle l\rangle^{-\delta}\,.
\end{equation}
The exponent $\psi$ counts the average number of parents per domain $\langle
p\rangle\equiv \sum p m_p/\sum m_p$:
\begin{equation}
\label{psi}
\langle p\rangle\sim \langle l\rangle^{1-\psi}.
\end{equation}
In the present case, $m_1=e^{-3t}$ and $\langle p\rangle=e^t$, while the
average domain length is $\langle l\rangle=e^{2t}$.  Therefore
\begin{equation}
\label{deltapsi}
\delta=\frac{3}{2},\qquad 
\psi=\frac{1}{2}.
\end{equation}
Even for the simplest models, these exponents are still known only
numerically, e.g., $\delta\approx 2.54$ and $\psi\approx 0.252$ for the
Ising-Glauber spin chain \cite{KB}, $\delta\approx 2.12$ and $\psi\approx
0.39$ for the Ising-Kawasaki spin chain \cite{BK}. For a few models, however,
the domain persistence exponents have been computed analytically
\cite{KB,LM,DLM}, e.g., for the random field Ising-Glauber spin chain
$\delta=\infty$ and $\psi=(3-\sqrt{5})/4=0.190983\ldots$ \cite{DLM}.

\subsection{Domain Length-Number Distribution}

The (normalized) domain length-number distribution $\rho_{lp}$ 
captures both the spatial and historical characteristics of the coarsening
domain mosaic and contains previous distributions: $\rho_l=\sum_p \rho_{lp}$
and $m_p=n\sum_l \rho_{lp}$. The domain length-number distribution satisfies
\begin{equation}
\label{rlp}
\frac{d\rho_{lp}}{dt}=-\rho_{lp}+
\sum_{i+j+k=l}\sum_{\alpha+\beta=p}\rho_{i\alpha} \rho_j \rho_{k\beta}. 
\end{equation}
To determine $\rho_{lp}$ we use the two-variable generating function
\begin{equation}
\label{cRdef}
{\cal R}(t,x,y)=\sum_{l\geq 1} \sum_{p\geq 1} x^l y^p\rho_{lp}(t).
\end{equation}
Multiplying Eq.~(\ref{rlp}) by $x^l y^p$, summing over all $l,p\geq 1$, and
using already known result (\ref{Rzt}) for the one-variable generating
function $\sum_j x^j\rho_j(t)$ we find that ${\cal R}$ satisfies
\begin{equation}
\label{cRt}
\frac{\partial {\cal R}}{\partial t}
=-{\cal R}+{\cal R}^2\,\frac{e^{-t}\,x}{\sqrt{1-(1-e^{-2t})x^2}}\,.
\end{equation}
Solving this equation subject to ${\cal R}(0,x,y)=xy$ we obtain
\begin{equation}
\label{cRsol}
{\cal R}(t,x,y)=\frac{e^{-t}\,xy}{1-y+y\sqrt{1-(1-e^{-2t})x^2}}\,.
\end{equation}
Expansion in $y$ is simple and for every $p$ we get 
\begin{equation}
\label{cRp}
\sum_{l\geq 1} x^l \rho_{lp}(t)=e^{-t}\,x
\left(1-\sqrt{1-(1-e^{-2t})x^2}\right)^{p-1}. 
\end{equation}
The expansion in $x$ is also straightforward and it leads to a series
representation for $\rho_{lp}$. Of course, $\rho_{lp}=0$ for even $l$'s. For
odd $l$'s we find 
\begin{eqnarray*}
\rho_{2l-1,p}&=&0 \qquad {\rm for}\quad l<p,\\
\rho_{2p-1,p}&=&e^{-t}\left(\frac{1-e^{-2t}}{2}\right)^{p-1},\\
\rho_{2p+1,p}&=&\frac{1}{2}\,(p-1)\,e^{-t}\left(\frac{1-e^{-2t}}{2}\right)^{p},\\
\rho_{2p+3,p}&=&\frac{1}{8}\,(p-1)(p+2)\,
e^{-t}\left(\frac{1-e^{-2t}}{2}\right)^{p+1},
\end{eqnarray*}
etc.  To extract the scaling behavior, it is more convenient to use
(\ref{cRsol}) rather than (\ref{cRp}). Writing
\begin{equation}
\label{xy}
x=1-e^{-2t}\xi, \qquad y=1-e^{-t}\eta,
\end{equation}
and taking the limit $t\to\infty$ we simplify (\ref{cRsol}) to
\begin{equation}
\label{cR1}
{\cal R}(\xi,\eta)=\frac{1}{\eta+\sqrt{1+2\xi}}\,.
\end{equation}
In the scaling limit $l,p,t\to\infty$ with the scaling variables
\begin{equation}
\label{scalLP}
L=l\, e^{-2t}={\rm finite},\quad
P=p\, e^{-t}={\rm finite},
\end{equation}
the domain length-number distribution admits the scaling form 
\begin{equation}
\label{rlpscal}
\rho_{lp}(t)=e^{-3t}{\cal H}(L,P).
\end{equation}
Inserting (\ref{xy}), (\ref{rlpscal}) into Eq.~(\ref{cRdef}) and replacing
summation by integration we convert the two-variable generating function into
the double Laplace transform
\begin{equation}
\label{cR2}
{\cal R}(\xi,\eta)=\int_0^\infty dL\,e^{-\xi L}
\int_0^\infty dP\,e^{-\eta P}\,{\cal H}(L,P)\,.
\end{equation}
Using Eqs.~(\ref{cR1}), (\ref{cR2}) and performing the inverse  Laplace
transform  we obtain 
\begin{equation}
\label{HLP}
{\cal H}(L,P)=\frac{P}{\sqrt{2\pi L^3}}\,
\exp\left(-\frac{L}{2}-\frac{P^2}{2L}\right)\,.
\end{equation}
Comparing (\ref{HLP}) with individual distributions (\ref{nscal}) and
(\ref{msol-scal}) we see that the domain length-number distribution does not
factorize even in the scaling limit.

{}From the length-number distribution one can extract a lot of things, e.g.
the fraction of persistent spins, i.e., spins which have not flipped during
the time interval $(0,t)$; this quantity usually decays as $\langle
l\rangle^{-\theta}$, where $\theta$ is the persistence exponent \cite{M}.
For the antiferromagnetic initial condition, for instance, the number of
persistent spins in a domain is exactly equal to the number of parents. The
average number of parents is
\begin{equation}
\label{pav}
\langle p\rangle_l=\frac{\sum_{p\geq 1}p\,\rho_{lp}}{\sum_{p\geq
    1}\rho_{lp}}\equiv \rho_l^{-1}\sum_{p\geq 1} p\,\rho_{lp}\,.
\end{equation}
In the long time limit, we can use (\ref{HLP}) and replace the summation by
integration. This leads to the asymptotic
\begin{equation}
\label{pav-asymp}
\langle p\rangle_l\to\sqrt{\frac{\pi l}{2}}\qquad {\rm when}\quad
l\to\infty. 
\end{equation}
The fraction of persistent spins $n^{-1}\sum_{l\geq 1}\langle
p\rangle_l\,n_l$ is now computed to give $e^{-t}=\langle l\rangle^{-1/2}$.
Thus, the persistence exponent is $\theta=1/2$.

\subsection{The $q$-state Potts model}

Some of the above calculations can be generalized to the case of the Potts
model.  For the $q$-state Potts model, each domain is in one of the $q$
possible states, and each time a domain is updated it adopts the state of one
of the two adjacent domains.  The updating of a domain results in merging of
all three domains with probability $1/(q-1)$ while with probability
$(q-2)/(q-1)$ only two domains merge.  Therefore the average number of
domains lost in every merging event is $2\times {1\over q-1}+{q-2\over
  q-1}={q\over q-1}$ leading to
\begin{equation}
\label{qnt}
\frac{dn}{dt}=-\frac{q}{q-1}\,n.
\end{equation}
Therefore the average domain size $\langle l\rangle=1/n$ increases
exponentially for arbitrary $q$. 

The number distribution $m_p(t)$ is also readily computable for the $q$-state
Potts model. One has
\begin{equation}
\label{qmpt}
\frac{dm_p}{dt}=-\frac{q+1}{q-1}\,m_p
+\frac{1}{q-1}\sum_{\alpha+\beta=p}\frac{m_\alpha m_\beta}{n}\,. 
\end{equation}
This equation admits the exponential ansatz (\ref{maa}) that reduces
(\ref{qmpt}) into a couple of differential equations
\begin{equation}
\label{qaa}
\frac{dA}{dt}=-\frac{q+1}{q-1}\,A,\qquad
\frac{da}{dt}=\frac{A}{(q-1)n},
\end{equation}
where $n=\exp\left[-\frac{q}{q-1}\,t\right]$.  Solving (\ref{qaa}) subject to
the initial conditions $A(0)=1$ and $a(0)=0$ (implied by
$m_p(0)=\delta_{p1}$) we obtain
\begin{equation}
\label{qmsol}
m_p(t)=Q^{q+1}(1-Q)^{p-1}, \quad
Q(t)\equiv e^{-t/(q-1)}.
\end{equation}
Re-expressing the quantities $m_1=Q^{q+1}$ and $\langle p\rangle=Q^{-1}$
through the average domain length $\langle l\rangle=n^{-1}=Q^{-q}$ we find
that the domain persistence exponents defined via equations
(\ref{delta})--(\ref{psi}) are given by
\begin{equation}
\label{qdp}
\delta=\frac{q+1}{q},\qquad 
\psi={q-1\over q}.
\end{equation}
The exponents thus obey the sum rule $\delta(q)+\psi(q)=2$.  

Now consider the length distribution. Particularly, the
normalized domain length distribution evolves according to 
\begin{equation}
\label{q-rho}
\frac{d\rho_l}{dt}=-\rho_l
+\frac{1}{q-1}\sum_{i+j+k=l}\rho_i \rho_j \rho_k 
+\frac{q-2}{q-1}\sum_{i+j=l}\rho_i \rho_j. 
\end{equation}
{}From this equation, we find an implicit relation for the generating
function (\ref{Rdef}) 
\begin{eqnarray*}
\frac{(1-R)^{q-1}\,(R+q-1)}{R^q}=
e^{qt}\,\frac{(1-R_0)^{q-1}\,(R_0+q-1)}{R_0^q}.
\end{eqnarray*}
This relation is a polynomial of $R$ of degree $q$. Hence it is impossible to
determine an explicit relation for the generating function, $R(t,x)={\cal
  F}[t,R_0(x)]$, apart from the Ising case ($q=2$) and two next cases
$q=3,4$.  The explicit expressions for $q=3,4$ are very involved so the exact
computation of $n_l(t)$ looks daunting.

Rather than seeking an exact solution, let's consider the asymptotic behavior. 
The scaling ansatz
\begin{equation}
\label{rho-scal}
\rho_l(t)=n\,{\cal F}(L),\qquad 
L=n\,l,
\end{equation}
recasts (\ref{q-rho}) into 
\begin{equation}
\label{***}
{\cal F}+qL\,\frac{d{\cal F}}{dL}+{\cal F}*{\cal F}*{\cal F}+(q-2){\cal F}*{\cal F}=0
\end{equation}
where symbol $*$ denotes the convolution operation, so that ${\cal F}*{\cal
  F}$ is the usual convolution $\int_0^L dL_1\,{\cal
  F}(L_1)\,{\cal F}(L-L_1)$, and ${\cal F}*{\cal F}*{\cal F}=\int_0^L
\int_0^L dL_1\,dL_2\,{\cal F}(L_1)\,{\cal F}(L_2){\cal F}(L-L_1-L_2)$.  The
Laplace transform $\Phi(s)=\int_0^\infty dL\,e^{-sL}\,{\cal F}(L)$ satisfies
\begin{equation}
\label{***-lap}
qs\,\frac{d\Phi}{ds}=\Phi^3+(q-2)\Phi^2-(q-1)\Phi,
\end{equation}
whose (implicit) solution reads
\begin{equation}
\label{Phi}
s=(1-\Phi)\left[\frac{\Phi+q-1}{q\Phi^q}\right]^{\frac{1}{q-1}}. 
\end{equation}
The sum rules $\sum \rho_l=1$ and 
$\sum l\,\rho_l=n^{-1}$ lead to 
\begin{equation}
\label{constraints}
\int_0^\infty  dL\,{\cal F}(L)=1, \qquad
\int_0^\infty  dL\,L\,{\cal F}(L)=1.  
\end{equation}
These two constraints determine first two constants in the small $s$
expansion of the Laplace transform: $\Phi(s)=1-s+\ldots$. This behavior was
taken into account in fixing a constant in the general solution to
Eq.~(\ref{***-lap}).

To complete the task, we must find $\Phi(s)$ from (\ref{Phi}) and then
perform the inverse Laplace transform. The first step therefore requires
finding a root of the polynomial of $\Phi$ of degree $q$. Thus it is
apparently impossible to find an explicit scaling solution when $q\geq 5$.
However, we can deduce the most interesting asymptotics for an arbitrary $q$.
For instance, Eq.~(\ref{Phi}) yields $\Phi\to (1-q^{-1})^{1/q}\,s^{-1+1/q}$
as $s\to\infty$, from which 
\begin{equation}
\label{small}
{\cal F}(L)\to \frac{\left(1-\frac{1}{q}\right)^{\frac{1}{q}}}
{\Gamma\left(1-\frac{1}{q}\right)}\,L^{-\frac{1}{q}}\qquad
{\rm as}\quad L\downarrow 0. 
\end{equation}
The large $L$ behavior of ${\cal F}(L)$ is reflected by the type of the
(closest to the origin) singularity of its Laplace transform. {}From
Eq.~(\ref{Phi}) we find that the singularity is the branch point located at 
$s_*=-q^{-1/(q-1)}$, namely
\begin{eqnarray*}
\Phi\to 2^{-\frac{1}{2}}\,q^{\frac{q-2}{2(q-1)}}
\left[s+q^{-\frac{1}{q-1}}\right]^{-\frac{1}{2}}\,,
\end{eqnarray*}
leading to 
\begin{equation}
\label{large}
{\cal F}(L)\to q^{\frac{q-2}{2(q-1)}}\,\frac{1}{\sqrt{2\pi L}}\,
\exp\left[-L\,q^{-\frac{1}{q-1}}\right]
\end{equation}
as $L\to \infty$. Thus the large $L$ behavior is qualitatively similar for
all $q$, while the small $L$ behavior is affected by the number of states. 

The Swendsen-Wang dynamics of the $q$-state Potts model is particularly
simple in the $q\to\infty$ limit when only two adjacent domains can merge.
The domain number distribution is trivial in this case,
$m_p(t)=e^{-t}\,\delta_{p1}$. The normalized domain length distribution
satisfies 
\begin{equation}
\label{inf-rho}
\frac{d\rho_l}{dt}=-\rho_l+
\sum_{i+j=l}\rho_i \rho_j. 
\end{equation}
This system of equations resembles (\ref{mpt}) and the solution is
accordingly found by seeking $\rho_l(t)$ in an exponential form like
(\ref{maa}). For initially uncorrelated Potts spins, all domains have initial
length one when $q=\infty$. Therefore $\rho_l(0)=\delta_{l1}$ and the
solution reads
\begin{equation}
\label{inf-sol}
\rho_l(t)=e^{-t}\left(1-e^{-t}\right)^{l-1}.
\end{equation}
The scaling form of the domain length distribution is exponential, ${\cal
  F}(L)=e^{-L}$; of course, this result can also be derived from (\ref{Phi})
which in the $q\to\infty$ limit simplifies to $\Phi=(1+s)^{-1}$.  An apparent
contradiction of the pure exponential scaled domain length distribution and
the general large $L$ asymptotic (\ref{large}) is the indication that the
limits $q\to\infty$ and $L\to\infty$ do not commute.

\section{Wolff Dynamics}

At zero temperature, the Wolff dynamics amounts for randomly choosing a spin
and flipping the whole domain containing that spin.  The flipping of a domain
$I$ again implies that it merges with its left and right neighboring domains
$I_-$ and $I_+$ to form a domain $I_-\cup I\cup I_+$.  In contrast with the
Swendsen-Wang dynamics, the flipping of a domain now occurs with rate
proportional to its length. Therefore the domain
density decreases with constant rate
\begin{equation}
\label{ntw}
\frac{dn}{dt}=-2,
\end{equation}
i.e., $n(t)=1-2t$ and the whole system reduces to a single domain at
$t_c=1/2$. 

The governing equations for the domain length densities read \cite{DH} 
\begin{equation}
\label{nltw}
{dn_l\over dt}=-ln_l-2{n_l\over n}+\sum_{i+j+k=l}\frac{jn_i n_j n_k}{n^2}\,. 
\end{equation}
We again use the normalized densities $\rho_l(t)$. They satisfy 
\begin{equation}
\label{rltw}
{d\rho_l\over dt}=-l\rho_l+\sum_{i+j+k=l}j\rho_i \rho_j \rho_k. 
\end{equation}
The generating function (\ref{Rdef}) satisfies 
\begin{equation}
\label{R}
{\partial R\over \partial t}=
-x{\partial R\over \partial x}+xR^2\,\frac{\partial R}{\partial x}.
\end{equation}
Changing variables from $(t,x)$ to $(T,X)\equiv (t,t-\ln x)$ 
removes the first term on the right-hand side of Eq.~(\ref{R}):
\begin{equation}
\label{RT}
\frac{\partial R}{\partial T}=-R^2\,\frac{\partial R}{\partial X}.
\end{equation}
Rewriting (\ref{RT}) for $X=T(R,X)$ gives
\begin{equation}
\label{X}
\frac{\partial X}{\partial T}=R^2
\end{equation}
which is solved to yield $X=F(R)+R^2T$, or 
\begin{equation}
\label{Xsol}
t-\ln x=F(R)+R^2 t,
\end{equation}
with $F(R)$ depending on initial conditions.  For the antiferromagnetic
initial condition $R_0(x)=x$ and thence $F(R)=-\ln R$, so the exact
solution (\ref{Xsol}) becomes
\begin{equation}
\label{xR}
x=R\,e^{t-R^2 t}\,.
\end{equation}
Clearly, Eq.~(\ref{xR}) gives an expansion of $x$ in terms of $R$.  We are
seeking the opposite, $R=\sum \rho_l x^l$. To determine $\rho_l$ we write
\begin{eqnarray*}
\rho_l=\frac{1}{2\pi \sqrt{-1}}\oint dx\,\frac{R(x)}{x^{l+1}}
      =\frac{1}{2\pi \sqrt{-1}}\oint dR\,\frac{R\, x'(R)}{[x(R)]^{l+1}},
\end{eqnarray*}
which is combined with Eq.~(\ref{xR}) to give
\begin{equation}
\label{lagrange}
\rho_l(t)=\frac{e^{-lt}}{2\pi \sqrt{-1}}\oint dR\,\left[\frac{e^{lR^2t}}{R^l}
-2t\,\frac{e^{lR^2t}}{R^{l-2}}\right].
\end{equation}
An elementary computation shows that $\rho_{2l}\equiv 0$ and 
\begin{equation}
\label{rafw}
\rho_{2l+1}(t)=\frac{(2l+1)^{l-1}}{l!}\,t^{l}\,\exp[-(2l+1)t]\,.
\end{equation}
Since $n_l=n\rho_l$ and $n=1-2t$, the densities vanish at $t_c=1/2$.  Note
also that in the proximity of the critical point (i.e., when $n\to 0$ and
$l\to\infty$), Eq.~(\ref{rafw}) simplifies to
\begin{equation}
\label{rnear}
\rho_l(t)\simeq \pi^{-1/2}\,l^{-3/2}\,\exp\left(-l n^2/4\right).
\end{equation}
This expression was previously derived in Ref.~\cite{DH} by a direct analysis
of the generating function near the critical point. Equation (\ref{rnear})
shows that $\langle l\rangle=n^{-1}$ does
not characterize the domain length distribution: Almost all domains are short
with lengths of order 1 but because the domain length distribution is
algebraic, $\rho_l\propto l^{-3/2}$ with a cutoff length of the order of
$n^{-2}$, the average domain length diverges as $n^{-1}$.

The single domain covers the entire spin chain at $t_c=1/2$, i.e., the system
undergoes a gelation transition. This transition differs from the ordinary
gelation transition that occurs in mean-field models of polymerization
\cite{pjf,hez} and evolving random graphs \cite{jlr} where the giant
component that is born at the critical time undergoes a long period of growth
before it engulfs the entire the system.  The difference from the ordinary
gelation transition is not surprising: In one dimension, the giant component
must cover the entire system, so the transition is discontinuous, while in
the mean-field models the transition is continuous. Despite of this important
physical distinction, the domain length distribution (\ref{rnear}) is very
similar to the cluster size distribution in polymerization \cite{hez} and the
component size distribution in evolving random graphs \cite{jlr}.

\section{Summary}

We demonstrated that the Ising chain endowed with zero-temperature
Swendsen-Wang dynamics exhibits scaling with the average length growing
exponentially with time.  We computed the domain length distribution in
special cases, e.g. for the antiferromagnetic initial condition and the
random initial condition. The scaled domain length distribution was shown to
be a product of a power-law and an exponential over the entire range of the
(scaled) length.  We also computed the domain number distribution, the domain
length-number distribution, and the domain persistence exponents.  The domain
length-number distribution does {\em not} factorize into the product of the
single variable distributions.  Some of the calculations have been
generalized to the Potts model; in particular, the domain number distribution
and the domain persistence exponents have been obtained.

We also studied the Ising chain endowed with zero-temperature Wolff dynamics
and showed that the system undergoes a kind of gelation -- below the critical
time the total number of domains is an extensive variable (proportional to
the system size) while at the critical time the entire system gets covered by
the single domain. This gelation transition is discontinuous since in one
dimension, the giant component must cover the entire system; in contrast, in
the mean-field models the gelation transition is continuous. Still we found
that the domain length distribution in the zero-temperature Ising-Wolff chain
is mathematically similar to the cluster size distribution in polymerization.

\ack I am thankful to Alan Sokal for useful remarks.

\end{document}